\documentclass[prl,floatfix,reprint]{revtex4}
\usepackage{graphicx}
\usepackage{subfigure}
\usepackage{amsmath}
\usepackage{epsfig}
\usepackage{epstopdf}
\usepackage{graphicx}
\usepackage{caption}
\usepackage{epstopdf}

\begin{document}


\title{Rich and Poor Cities in Europe. An Urban Scaling Approach to Mapping European Economic Transitions}

\author{Emanuele Strano}

\affiliation{Department of Civil and Environmental Engineering, Massachusetts Institute of Technology (MIT), Cambridge, MA 02139, USA}
\affiliation{Department of Civil and Environmental Engineering, Polytechnic School Of Lausanne (EPFL), Lausanne , CH}
\affiliation{emanuele.strano@epfl.ch}

\author{Vishal Sood}
\affiliation{Department of Civil and Environmental Engineering, Polytechnic School Of Lausanne (EPFL),Lausanne , CH}

\maketitle

\date{\today}

\section{Abstract}
Recent advances in the urban science make broad use of the notion of scaling. We focus here on the important scaling relationship between the gross metropolitan product ($GMP$) of a city and its population ($pop$). It has been demonstrated that  $GMP \propto Y pop ^{\beta}$ with $\beta$ always greater than 1 and close to $1.2$. This fundamental finding highlights a universal rule that holds across countries and cultures and might explain the very nature of cities. However, in an increasingly connected world, the hypothesis that the economy of a city solely depends on its population might be questionable. Using data for 248 cities in the European Union between 2005 and 2010, we found a double $GMP/pop$ scaling regime. For West EU cities, $\beta = 1$ over the whole the period, while for post-communist cities $\beta >1$ and increases from $\sim1.2$ to $\sim1.4$. 
The evolution of the scaling exponent describes the convergence of post-communist European cities to open and liberal economies. We propose a simple model of economic convergence in which, under stable political conditions, a linear $GMP/pop$ scaling is expected for all cities. The results suggest that the $GMP/pop$ super-linear scaling represents a phase of economic growth rather than a steady, universal urban feature.  
The results also suggest that relationships between cities are embedded in their political and economic context and cannot be neglected in explanations of cities, urbanization and urban economics.

\section{Introduction}
The hypothesis of urban scaling is certainly a pillar of quantitative methods for understanding cities \cite{Batty2008}. 
Far from neglecting the uniqueness of each urban settlement, the urban scaling hypothesis highlights common patterns among dissimilar cities. In a way, all cities are considered \emph{variations on a theme} \cite{Bett1}, which are ultimately governed by universal and simple laws that regulate their form and functions. The meaning of scaling for cities lies in acknowledging their bottom-up and emergent nature rather than any top-down regulation that might come from urban planning, urban policies or local socio-economic conditions \cite{Jacobs1961}. We focus on the scaling laws that regulate the economic output of a city, or gross metropolitan product ($GMP$), in relation to its population ($pop$). Any geographically extended social entity's per capita gross domestic product (GDP) is expected to be independent of its population. After all, this is the mean economic output of the people living in that city, village or metropolis. A population-independent per capita GDP would imply a linear relationship between total GDP and population.
However, urban historians and urban economy scholars \cite{Jacobs1970, Romer1986, Glaeser1991} have noted a special behaviour of cities' economic performance 
Concepts such as \emph{dynamic externalities} and \emph{urbanization economies} are rooted in a combination of urban density and diversity, which promotes both interactions (knowledge exchange) and economic competition that in turn promotes innovation and economic growth. These observations focused on urban vs rural economies, highlighting the primal role of cities in economic growth and fostering further understanding of urban economic output resulting from individuals' interactions in dense urban environments. Important investigations have been proposed in this direction by Bettencourt, Lobo and colleagues \cite{Bettencourt2007, Bettencourt2013, Bett1}. Inspired by allometric laws, which regulate the relationships between body mass and physiology in animals and plants \cite{Kleiber47, West1997}, they propose a unified theory of urban systems \cite{Bettencourt2010}. For example, they observe that  $GMP$ is universally dependent on $pop$ through a power law such as $GMP \propto Y pop ^{\beta}$ with $\beta$ always greater than 1 and close to $1.2 \pm 0.02$. Larger cities are wealthier than expected by a linear assumption when compared with smaller cities. Super-linear $GMP/pop$ scaling ($\beta>1$) provides a strong and simple quantitative explanation for the increasing growth and expansion of cities: large urban basins offer more opportunities than disadvantages compared to smaller basins, which is why they continue to grow and attract people. Super-linear scaling is consistent with classical urban economic theories. Such theories are built upon the established fact that urbanization in developed countries was accompanied by economic growth and industrialization. These historic patterns generated expectations of a virtuous circle between economic growth and urbanization, regardless of the local conditions \cite{Spence2009, Duranton2014}. However, despite its great importance, the universality of super-linear $GMP/pop$ scaling is still debated and poses some important questions about the relationships between urbanization and economic growth. For example, recent geographically sound studies show that the definition of urban boundaries is crucial to the measurement of any scaling of socio-environmental performance with population \cite{Makse2014,Arcaute2013}. 
As far as geography matters, they make a reasonable point. However, by focusing on the effects of the spatial definition of a city, they implicitly accept that scaling occurs inside a metropolitan boundary, which is not far from the core urban scaling assumption. More, and perhaps more important, questions related to the increasing urbanization rates of persistently poor, non-industrialized countries are raised by recent studies \cite{Gollin2013, Glaeser2014}.  This observation is not consistent with the assumption that urban economic growth is determined solely by its citizens' interactions within urban limits. Some natural questions arise when cities are not considered as closed and independent systems: Why, given the same rate of urbanization, does Asia contain a number of the most explosive economies while sub-Saharan Africa has seen very little growth? How can we explain the stagnant economic conditions of a metropolis such as Dhaka? What is the meaning of urban scaling in rapidly growing emerging economies? Finally, is it true that all cities, regardless of their economic environments and histories, are \emph{variations on a theme}? 

We explored these questions using archival data on per capita GDP and population for 248 cities in the European Union between 2005 and 2010. We failed to verify the consensus that $GMP$ is universally super-linear with population. European metropolitan areas do not show a uniform and super-linear behaviour in terms of GMP scaling. Former Eastern Bloc cities behave differently than other cities in Europe, suggesting that positive externalities are not sufficient for determining economic growth. We found, however, that different scaling regimes mirror the economic transitions and economic convergence patterns of post-communist and former Eastern Bloc cities. 
The results suggest that super-linear $GMP/pop$ scaling represents a transition condition of growing economies rather then a universal and stable condition of all cities. Moreover, the economic growth of a town is dependent not only on its population size but indeed on higher level economic and social dynamics.

\section{Data and methods}
We used GMP and population for 243 metropolitan areas in Europe over the decade 2005-2010. The data are provided by Eurostat  \cite{UrbanAtlas2}. These data are public and freely available from the Eurostat website \cite{UrbanAtlas2}. The analysis of the scaling relationship was performed by linear regression after log transformation of the variables. The definition of a city's boundary is crucial because it defines the unit or scale of analysis. Because of the intrinsic spatial complexity of cities and urban spatial contiguity due to conurbation phenomena, theories of the spatial and political definition of cities as units of analysis are controversial \cite{Batty2011}. Different approaches to solving this problem have been proposed \cite{Arcaute2013, Makse2014}.  It is worth noting that super-linear scaling of GMP/pop size has been proposed mostly for metropolitan statistical areas (MSA) in the United States, which are essentially large urbanized areas within the same labour market basin and are typically defined by a large, dense urban core including small satellite cities and conurbation land \cite{Bureau2008}. In this study, we used the European equivalent of the MSA, the Large Urban Zone (LUZ), as defined by the European statistic office \cite{UrbanAtlas2}. The  similar nature of MSAs and LUZs ensures that our results are consistent with the previous literature.

\section{Results: Linear and super-linear GMP/pop scaling mirrors convergent economies in European Union}
\begin{figure}[h!]
\begin{center}
\includegraphics[scale=0.4]{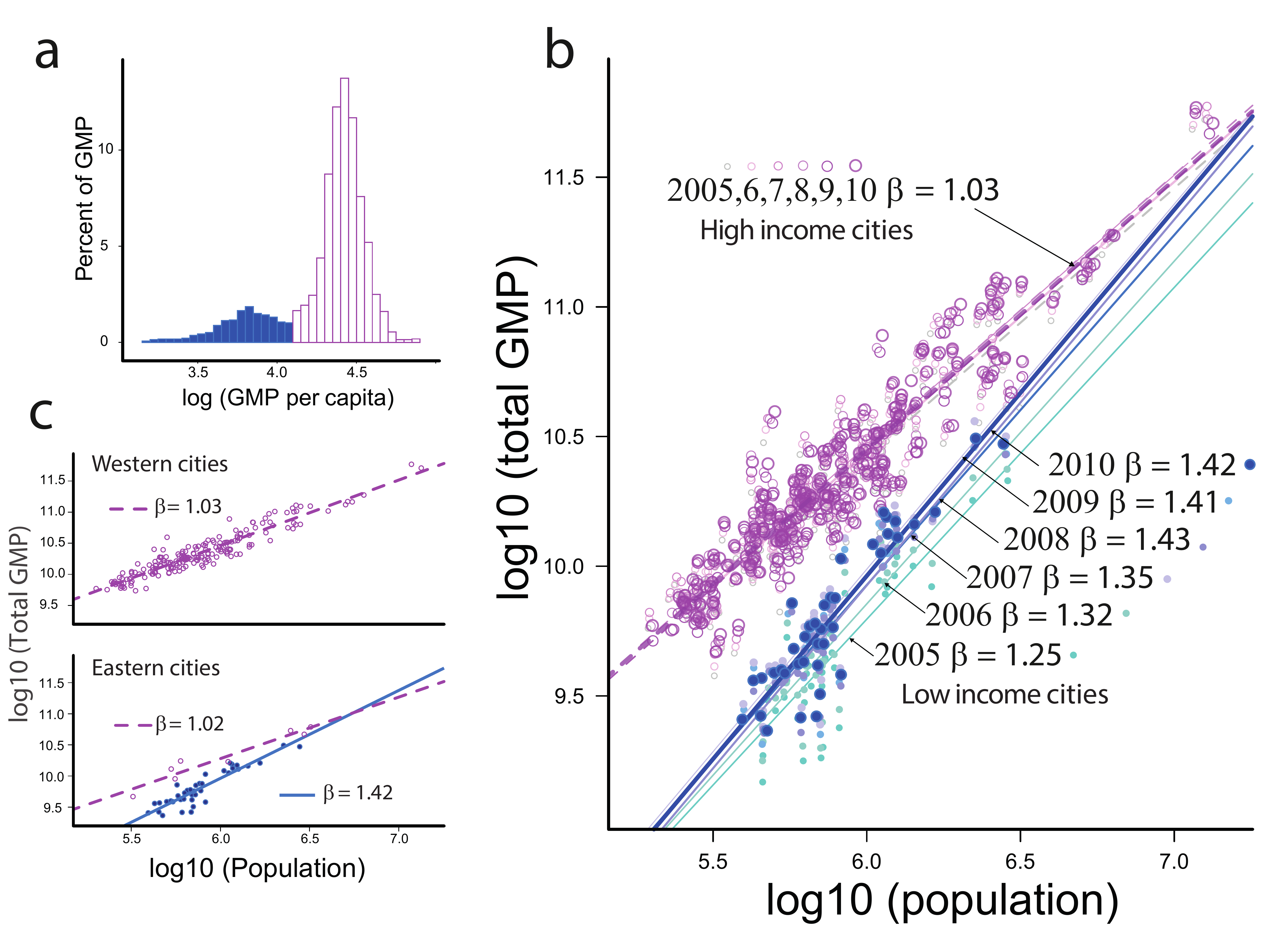}
\caption { a) Histogram of the per capita $GMP$ of the 248 cities. The double bell indicating two classes of economies is evident. We called these low-income (filled blue) and high-income (empty purple) groups. 
Using the same colour code, b) shows the $GMP/pop$ scaling for low- and high-income cities from 2005 to 2010. High-income cities have a stable linear exponent $\beta = 1.03 \pm 0.05 $ over the entire period. Low-income cities show super-linear scaling with an increasing trend from $\beta = 1.25 \pm 0.25 $ in 2005 to $\beta = 1.42 \pm 0.20 $ in 2010. c) Shows $GMP/pop$ in 2010 for the Eastern Bloc (bottom panel) and the Western Bloc (upper panel). It is possible to observe that no cities with low incomes are in the Western Bloc, while few cities with high incomes are in the Eastern Bloc.}
\label{fig_1}
\end{center}
\end{figure}

The Europe Union is composed of two major groups of countries, Western European countries (WEu) and former members of the Eastern Bloc (EEu). The EEu countries were aligned with the USSR before its collapse in the late 1980s. EEu cities experienced transitions from totalitarian to liberal economic regimes. Theories of the dynamics of transition from totalitarian to liberal economies are widely debated \cite{Svejnar2002}. The main expectation for transition economies is fast economic growth due to market liberalization, increased private and foreign investments and an emerging private banking system. 
Given the low per-capita income, transition economies can grow faster than richer economies. Transition economies may eventually converge to developed economies.   
WEu and EEu perfectly fit the picture of convergent economies.
The distribution of per capita $GMP$ of the 248 cities is bimodal (Fig.\ref{fig_1}a), indicating the existence of two groups, low- and high-income cities.
The partition threshold dividing low- from high-income cities is 11,400 euros. With few exceptions, as shown in Fig.\ref{fig_1}c, the two groups overlap with the East-West European partition.
WEu cities are richer than EEu cities. However, we would not expect the double peaked wealth to affect the super-linear $GMP/pop$ scaling. However, EEu and WEu cities have very different scaling regimes with population. The exponent $\beta$ of $GMP$ on population for the EEu countries is $>1$ and always larger than that of WEu cities. The value increases over time from $\beta = 1.25 \pm 0.25$ in 2005 to $\beta = 1.42 \pm 0.20 $ in 2010. On the other hand, in WEu cities, $\beta = 1 \pm 0.05$ over the entire period; these results are reported in Fig.\ref{fig_1}b.

We propose a simple descriptive model of the observed double scaling regime that might mirror the process of economic convergence of EEu to WEu cities incorporating the diminishing return effect. The scaling exponent of $GMP$ versus population increases, which indicates that the $GMP$ growth rate is proportional to its population. Indeed, if a city with a larger population grows faster than a city with a smaller population, its $GMP$ vs population curve (on a log scale) will become steeper over time, which this is what we observe for the emerging EEu economies. 
Given a city $i$ with log population $p_i$, let $g_{i, t}$ be its log GMP for year $t$.  We assume at this stage that the population of cities does not change to the same degree as its GMP. This assumption works well with the observed data. For our analysis, we used a log scale for GMP ($g$) and population size ($p$). On a log scale, the scaling of GMP versus population size can be given a mathematical form:
\begin{equation} \label{gmp_pop}
 g_{i, t} = \alpha_t + \beta_t p_i
\end{equation}
where the (time-dependent) exponent $\beta_t$ indicates whether the relationship is linear ($\beta = 1$), sub-linear ($\beta<1$) or super-linear ($\beta>1$). Empirically, we observe that $\beta_t$ is 1 for WEu cities for all years and increasingly larger than 1 for EEu cities. The growth of $GMP$ can be modelled as a change from year to year:
\begin{equation} \label{gmp_growth_1}
g_{i, t+1} = g_{i, t} + \gamma_0 + \gamma_1 p_i.
\end{equation}
Economic models of $GMP$ growth have assumed a constant, population-independent $GMP$ growth rate: 

\begin{equation} \label{gmp_growth_2}
g_{i, t+1} = g_{i, t} + \gamma_0,
\end{equation}

with $\gamma_1 = 0$. Plugging the $GMP-pop$ scaling relationship into the $GMP$ growth model, we find the relationship for the change in the scaling exponent:
\[
  \alpha_{t+1} + \beta_{t+1} p_i = \alpha_{t} + \gamma_i + (\beta_{t} +  \gamma_1) p_i.
\]
Equating the population-dependent terms on both sides of the equation we find:
\begin{equation} \label{beta_growth}
\beta_{t+1} = \beta_t + \gamma_1.
\end{equation}

Population-independent $GMP$ growth (where $\gamma_1 = 0$) would lead to a stable scaling exponent over time, which we observe for WEu cities. For EEu cities, we observe a $\beta$ that grows for cities that are poorer than the West and a $\beta$ that is equal to 1 and stable for those that are as rich as cities in the West. To accommodate this latter observation, we make the $GMP$ growth rate depend on $GMP$ itself:
\begin{equation} \label{gmp_dep_growth}
g_{i, t+1} = g_{i, t} + \gamma_0 + \frac{\gamma_1 p_i}{1 + exp(\theta(g_{i, t} - p_i - g_{\star}))}
\end{equation}

Here, the population size-dependent term in the growth rate also depends on the city's $GMP$ such that this term is smaller when the per capita $GMP$ of the city is larger than a threshold value of $g_{\star}$. The value of $g_{\star}$ is indeed arbitrary and can change given specific conditions. In our case, $g_{\star} = 11400 ~euros$.
Note that $g_{i, t} - p_i$ is {\bf log per capita GMP}.
This new form of $GMP$ growth reduces to Eq.(\ref{gmp_growth_1}) for cities that have a small $GMP$ ($ g_{i, t} - p_i \ll g_{\star}$). Cities keep growing at a population- and $GMP$-dependent rate until their $per capita GMP$ crosses $g_{\star}$. When the city is rich, as is the case for all cities in the West and a handful in the East, the population- and $GMP$-dependent growth term drops out. Over very long periods and under desirable, stable economic and political conditions, we would expect all cities to cross the wealth threshold, and the growth term becomes constant and independent of the population. In the limit of $\theta \to \infty$, the population-dependent growth term in Eq.(\ref{gmp_dep_growth}) drops out as soon as $g_{i,t} - p_i = g_{\star}$. Thereafter, $GMP$ growth follows Eq.(\ref{gmp_growth_2}), increasing at a constant 
{\it population-independent} rate $\gamma_0$, starting at a value $g_{\star} +
p_i$ and thus scaling linearly with $p_i$.
After a very long time, we expect all cities' $GMP$ to show linear scaling with population. This apparently simple relationship is prone to speculation. 

In our opinion, linear scaling ($\beta = 1$), as observed for WEu cities, may reflect the economically mature and politically stable conditions of the European Union. Linear scaling may be also a consequence of European redistribution of incomes to reduce disparities between small and large cities. Recent investments in internet connectivity and international high-speed trains have consequently increased the proximity of European capital cities, and for WEu economic activity, an average individual may be able to increasingly interact with out-of-town economic agents. 
As a result, the GMPs of different cities can become too inextricably interconnected to satisfy the assumption of independent LUZs (or MSAs). In the limit where an individual can interact with anyone in the larger economy of a state, his/her economic output will depend on the total population of the larger economic basin (here, the entire EU) and is thus independent of the population of the city where she/he resides. This leads to the traditional linear GDP vs population size relationship. The linearity of the $GMP-pop$ relation as a signature of mature economies is also indicated by the exponents of the richer Eastern cities. The exponents for these cities are closer to 1 than are those for the Eastern Bloc as a whole (Fig.\ref{fig_1}c). Thus, the exact scaling of $GMP$ with population depends not only on the geographic/historical context but also (an perhaps solely) on the wealth of the city.
The situation in Eastern cities tells a different story. After the collapse of the USSR, these economies developed rapidly, and more populous cities have been developing increasingly quickly over time. As expected, more populous cities have seen higher rates of growth, causing the exponent $\beta$ to increase from 1.25 to 1.42. An increasing $\beta$ indicates improving economic performance but may also indicate increasing inequality between cities. In a way, the observed hyper-linear scaling reflects an unbalanced situation of rapid growth of large cities and economic segregation of small ones, which makes the redistribution of income increasingly difficult. This finding is consistent with recent observations about the economic conditions and increasing economic inequality of Eastern European countries \cite{Heyns2005}. Increasing super-linearity can also indicate a  delay in economic convergence between EEu and WEu economies.

\section{Conclusions}

Our results might be relevant to two main debates on i) the universality of scaling laws for cities and, consequently, on ii) on consistence and dependences between different types of urbanization economies. 

The hypothesis of urban scaling certainly represents a large step in the scientific understanding of cities. We have shown that $GMP/pop$ scaling is a powerful tool for observing economic fluctuations and convergences even over shorter periods. However, the universality of super-linear $GMP/pop$ scaling remains difficult to prove. This in turn suggests that increasing connections among cities, rather than solely within cities, play crucial roles in urban economies and cannot be neglected in understandings urban economic and social dynamics. It might be possible also that super-linear scaling is an effect of aggregating many cities from different locations in the same analysis. 
In general terms, one might criticize the application quantitative tools inspired by physics and the natural sciences to cities, which reduces cities to natural or physical systems. Such systems (of cities) must then follow universal laws as any gas must expand with higher temperatures. The appeal of human/nature metaphors and of physical determinism in the study of cities, as for any simple and universal solution, is unquestionable. The simple results presented in this paper suggests, however, that very existence of universal laws that govern cities is questionable. Cities and urban systems can be indeed described using quantitative methodologies as with physical systems, but this does not justify the equation of cities to pure physical systems.
In this sense is important to highlight that economic transition, specially in the case of Europe, is a top-down and planned phenomena and it played a crucial role in the global economy framework.  
On the other hand, cities are highly complex and self-organizing systems and we cannot neglect this complexity.
In our opinion, a balance between quantitative and qualitative analysis, as in the proposed study, is key to extracting information from the data and to producing a more objective understanding of cities.  

Other observations arise from observing the coexistence of distinct economic groups within the EU' s cities. 
The causes of this separation are certainly rooted in the history of Europe. The causes of economic convergence are due to the transition from totalitarianism to liberal economies experienced by the Eastern countries. Certainly, as for the well-known Baltic Tigers, the advantageous conditions of Western states and extra-EU investors trigger a virtuous cycle of out-of-town economic input, consequently contributing to the observed increase in the $ GMP/pop$ scaling exponent.
Other factors, such as monetary changes or economic regulations, played a role in creating this separation and they should be taken in consideration for such analysis.
Regardless of the causes of economic separation and their dynamics, it is interesting to observe that the fast  economic convergence increases inequalities between cities in developing countries. 
It is possible that fast economic convergence due to forced liberalization may trigger further separation of poor and rich cities. Coexistence of poor and rich cities in a unique monetary and economic system might also suggest the existence of different urbanization economies which are rooted not only in economic growth but on different, and maybe hidden, urbanization drivers. Considering that most of the next urbanization is taking place in developing countries, such hidden and non GMP-based urbanization drivers must be further investigate.

%


\section{Acknowledgements}
The authors thank Jos\'e Lobo for his valuable support and advice.
They also thank Karen C. Seto, Matthew Parkan and Andrea Rinaldo for their suggestions.

\end{document}